\documentclass[smallextended]{svjour3}       % onecolumn (second format)

\smartqed  % flush right qed marks, e.g. at end of proof

\usepackage{graphicx}

\usepackage{mathptmx}      % use Times fonts if available on your TeX system

\usepackage{amsmath,amssymb,amsfonts,mathtools}
\usepackage{tikz}
\usetikzlibrary{decorations.markings}

\newcommand{\dd}{\mathrm{d}}

\usepackage{hyperref}

\begin{document}

\title{Scales and hierachies in \\ asymptotically safe quantum gravity: a review
%	\footnote{Prepared for the proceedings of the Workshop ``Naturalness, Hierarchy, and Fine Tuning'', Aachen, 28.2.-2.3.2018.}
}

\author{ Frank Saueressig \and Giulia Gubitosi \and Chris Ripken }

\institute{ F. Saueressig \at
	Institute of Mathematics, Astrophysics and Particle Physics (IMAPP), Radboud University \\
	\email{f.saueressig@science.ru.nl} 
	\and
	G. Gubitosi \at
              Institute for Theoretical Physics, University of Wroclaw \\
              \email{giulia.gubitosi@gmail.com}           %  \\
           \and
           C. Ripken \at
              Institute of Mathematics, Astrophysics and Particle Physics (IMAPP), Radboud University \\
              \email{aripken@science.ru.nl}           %  \\
                    %  \\
}


\maketitle

%------------------------------------------------------------------
\begin{abstract}
%------------------------------------------------------------------	
The asymptotic safety program strives for a consistent description of gravity as a non-perturbatively renormalizable quantum field theory. In this framework the gravitational interactions are encoded in a renormalization group flow connecting the quantum gravity regime at trans-Planckian scales to observable low-energy physics. Our proceedings reviews the key elements underlying the predictive power of the construction and summarizes the state-of-the-art in determining its free parameters. The explicit construction of a realistic renormalization group trajectory describing our world shows that the flow dynamically generates two scales: the Planck scale where Newton's coupling becomes constant and a terrestrial scale where the cosmological constant freezes out. We also review the perspectives of determining the free parameters of the theory through cosmological observations.
%------------------------------------------------------------------
\keywords{Quantum Gravity, Renormalization Group, Planck Scale, Cosmological Predictions}
\end{abstract}
%-----------------------------------------------------------------
\section{Introduction}
One of the key challenges in any quantum gravity program is
to explain, or at least accommodate, the tiny value of
the cosmological constant found in cosmological observations \cite{PDG2018}.
From the quantum perspective, the cosmological constant problem (see, e.g., \cite{Weinberg:1988cp,Martin:2012bt,Sola:2013gha,Padilla:2015aaa}
and references therein) is often considered as the biggest mismatch between theoretical expectations and experimental observations throughout physics. Summing up the vacuum contributions in a field theory with a ultraviolet (UV) cutoff $\Lambda_{\rm UV}$ one expects that the value of the cosmological constant is given by
\begin{equation}\label{LambdaUVcutoff}
\Lambda \propto \Lambda_{\rm UV}^2
\end{equation}
with the constant of proportionality being of order one. If $\Lambda_{\rm UV}$ is identified with the Planck scale
\begin{equation}\label{MPlanck}
M_{\rm Pl}
=
\sqrt{\frac{\hbar c}{8	\pi G}}
=
2.4	\times	10^{27}	\text{ eV}
\text{,}
\end{equation}
this implies that the ``natural'' value of the cosmological constant $\Lambda \propto M_{\rm Pl}^2$ should be linked to the strength of the gravitational interactions given by Newton's constant $G$. Cosmological observations \cite{PDG2018} indicate that $\Lambda$ is much smaller though
\begin{equation}\label{Lambdaobs}
\Lambda_{\rm obs}
=
4	\times	10^{-66}	\text{ eV}^2
\simeq
10^{-120} \, M_{\rm Pl}^2
\text{.}
\end{equation}
Thus the observed value dwarfs the theoretical expectation by 120 orders of magnitude.\footnote{The cosmological constant problem also persists if one just considers the vacuum energy contributed by the electroweak symmetry breaking \cite{Martin:2012bt}, $\Lambda_{\rm ew} = - 10^{55} \Lambda_{\rm obs}$, even though at a slightly less severe level. Since we will consider the case of pure gravity only, we will not discuss contributions originating from the matter sector in the sequel.} On the theoretical side this may be accommodated by introducing a bare value $\Lambda_{\rm bare}$ at the Planck scale which then cancels the contributions from the field modes. In order to match with \eqref{Lambdaobs} these two contributions must cancel to an accuracy of 120 digits. It is then very hard to envision a mechanism where the various contributions to the vacuum energy are finetuned in such a way that they lead to the observed value. From this point $\Lambda_{\rm obs}$ is considered completely unnatural. One may hope that a quantum theory of the gravitational interactions may shed some light on this puzzle.

In this proceedings we review the status of the cosmological constant within one particular approach to quantum gravity, the asymptotic safety program \cite{Niedermaier:2006wt,Codello:2008vh,Litim:2011cp,Reuter:2012id,Nagy:2012ef} (also see \cite{PercacciBook,RSBook} for recent textbooks and \cite{Eichhorn:2018yfc} for an overview on asymptotically safe gravity matter systems). A key difference to the effective field theory framework, where the theory is considered to be valid below a certain UV cutoff $\Lambda_{\rm UV}$ only, is that asymptotic safety ensures that the construction remains consistent up to arbitrarily short length or, equivalently, arbitrarily high energy scales. This entails in particular that there is no UV cutoff which could naturally appear in the relation \eqref{LambdaUVcutoff}. 

Technically, the high-energy regime of the gravitational interactions is controlled by a fixed point of the underlying renormalization group flow.
Besides leading to a quantum field theory valid at the highest energy scales, this also provides predictive power in the sense that not all candidate theories for quantum gravity will approach this renormalization group fixed point in the UV. The condition that they do fixes an infinite number of gravitational couplings in terms of a small number of free parameters. Conceptually, $\Lambda_{\rm obs}$ should then be considered as part of the experimental input which must be taken from observation to identify the quantum gravity theory realized by Nature.

The large hierarchy between the Planck scale and the cosmological constant then reflects itself in the energy dependence of the couplings (see Fig.\ \ref{fig:einsteinhilbert_phystraj}). Newton's coupling $G$ freezes out at the Planck scale thereby generating the scale $k_G$ dynamically. The small value of $\Lambda_{\rm obs}$ generates a second scale $k_\Lambda \ll k_G$. For energies $k < k_\Lambda$ the cosmological constant is indeed constant while in the intermediate regime $k_\Lambda < k < k_G$ one finds $\Lambda \propto k^4$. The transition scale $k_\Lambda$ is set by $\Lambda_{\rm obs}$ and for $\Lambda_{\rm obs} = 0$ would be given by $k_{\Lambda} = 0$.  

Besides the observed values of Newton's coupling and the cosmological constant, it is conceivable that asymptotic safety possesses additional free parameters. Typically, these are associated with higher-derivative (HD) interactions, as e.g.,
\begin{equation}\label{HDaction}
S^{\rm HD} = \frac{B}{16 \pi G} \int \dd^4x \sqrt{g} \, R^2 \, . 
\end{equation}
Interactions of this type are notoriously difficult to observe, e.g., at solar system scales \cite{Will:2014kxa}. Nevertheless, the underlying modified gravitational dynamics may have left imprints during the very early stages of the cosmic evolution which may still be visible in the sky today. Thus, trying to explain some observable features based on modified dynamics of gravity may allow to find values for such couplings as well.

The rest of this review is organized as follows. Section \ref{sec:RGandAS} summarizes the key concepts underlying 
 asymptotic safety together with the renormalization group techniques used to explore this scenario in the context of gravity.
 In Section \ref{sec:EinsteinHilbert} we discuss an approximation of the renormalization group trajectory realized in Nature based on the Einstein-Hilbert action before improving it by including higher-order scalar curvature terms. The prospects of fixing the free parameters of the theory based on cosmological observations are discussed in Section  \ref{sec:cosmology} and  some concluding remarks are given in Section \ref{sec:conclusion}. 

%------------------------------------------------------------------------------------------
\section{Renormalization group and Asymptotic Safety}\label{sec:RGandAS}
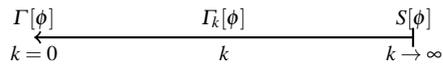
\begin{figure}
	\centering
	\begin{tikzpicture}
	\draw [<-|,thick] (0,0) 
	node[below] {$k=0$} 
	node[above]	{$\Gamma[\phi]$}
	--	(2.5,0)
	node[below] {$k$}
	node[above]	{$\Gamma_k[\phi]$}
	-- (5,0) 
	node[below] {$k \to \infty$}
	node[above] {$S[\phi]$}	;
	\end{tikzpicture}
	
	\caption{Schematic overview of the renormalization group. The effective average action $\Gamma_k$ interpolates between the quantum effective action $\Gamma \equiv \Gamma_0$ in the IR and the bare action $S$ in the UV.}
	\label{fig:renormalizationgroup}
\end{figure}

%------------------------------------------------------------------------------------------
The investigation of Asymptotic Safety is closely linked to the 
 key idea of the Wilsonian renormalization group (RG) where  quantum fluctuations are ``integrated out'' consecutively, shell-by-shell in momentum space. The scale-dependent dynamics at the scale $k$ is then captured by the effective average action (EAA) functional $\Gamma_k[\phi]$ \cite{Wetterich:1992yh,Morris:1993qb,Reuter:1993kw,Reuter:1996cp}
whose effective interactions contain all quantum corrections from fluctuations with momenta $p^2 \gtrsim k^2$. The flow of $\Gamma_k$ with respect to the RG parameter $k$  connects physics at different energy scales. When $k\to \infty$, no quantum fluctuations have been integrated out, and $\Gamma_k$ essentially reduces to the ``bare action'' $S[\phi]$. Since this occurs at high energies, this regime will be referred to as the UV. On the other hand, as $k\to 0$, all quantum fluctuations have been taken into account. In that case, $\Gamma_k$ reduces to the full quantum effective action $\Gamma$. For obvious reasons, this will be referred to as the infrared (IR) regime. See also Fig.\ \ref{fig:renormalizationgroup}.

In order to study the flow of the EAA, the action functional $\Gamma_k$ is expanded in a suitable operator basis $\{\mathcal{O}_i[\phi]\}$ containing all interaction monomials compatible with the symmetries of the theory
\begin{equation}\label{eq:operatorexpansion}
\Gamma_k[\phi]
=
\sum_i	u_i(k)	\mathcal{O}_i[\phi] \, . 
\end{equation}
The coefficients $u_i(k)$ are the coordinates of $\Gamma_k$ with respect to this basis. 

Introducing the logarithmic RG scale  $t = \log (k/k_0)$, with an arbitrary reference scale $k_0$,
the scaling of the theory is then captured by the $\beta$-functions of the coupling constants
\begin{equation}\label{beta1}
\partial_t	u_i(k)
=
\beta_{u_i}( \{ u_j	\} ) \, . 
\end{equation}
 The $\beta$-functions can be calculated, for instance in perturbation theory \cite{Weinberg:1996kr} or by functional methods \cite{Polchinski:1983gv,Wetterich:1992yh}. Solutions of the set of equations \eqref{beta1} are called RG trajectories.

In order to obtain a theory that is healthy, the couplings $u_i$ should remain finite at all scales. Furthermore, in order to be able to make the theory predictive, only finitely many couplings should be measured to characterise the entire RG flow. A theory satisfying both conditions is referred to as renormalizable.

The latter condition may prove to be problematic in the light of the infinite sum in Eq. \eqref{eq:operatorexpansion}. However, the other condition actually gives a way out to this problem. Since the couplings are to remain finite at all scales, the RG trajectory has to have an endpoint where the couplings do not change anymore. At this point, all $\beta$-functions vanish simultaneously
\begin{equation}
\left. \beta_{u_i} \right|_{u_j = u_j^*} = 0 \, , \qquad \forall \, i . 
\end{equation}
The point $\{u_j^*\}$ is therefore a fixed point of the RG flow. If the fixed point occurs at the point where all interactions are turned off, we speak of a Gaussian fixed point (GFP). Theories attracted to a GFP at high energies are termed asymptotically free.  If the fixed point contains interactions, one refers to a non-Gaussian fixed point (NGFP) and theories approaching the NGFP at high energies are called asymptotically safe.

\begin{figure}
	\centering
	\includegraphics[width=.5\textwidth]{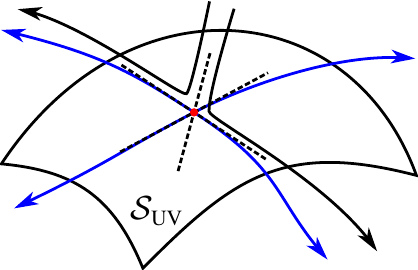}
	\caption{Illustration of the UV-critical hypersurface $\mathcal{S}_{\rm UV}$ associated to a renormalization group fixed point. The fixed point is denoted in red. Trajectories that end in the UV at the fixed point are represented by blue lines; arrows point from UV to IR. Together, the trajectories connected to the fixed point span the UV-critical hypersurface. By definition, trajectories that are not contained in $\mathcal{S}_{\rm UV}$ are eventually driven away from the fixed point. The eigendirections of the linearized flow are denoted by the dashed lines.}
	\label{fig:uvcritical}
\end{figure}

The requirement that the RG flow has a fixed point in the UV also solves the problem of predictivity. To this end, consider the RG trajectories that end up in a UV fixed point. This set spans the UV-critical hypersurface $\mathcal{S}_{\rm UV}$ embedded in the space of actions spanned by the $\mathcal{O}_i$; see also Fig.\  \ref{fig:uvcritical}. If this hypersurface is finite-dimensional, one requires only finitely many couplings in order to specify a particular RG trajectory within $\mathcal{S}_{\rm UV}$. The condition that gravity should be described by a trajectory within $\mathcal{S}_{\rm UV}$ restores predictivity of the construction. In a slight abuse of string theory nomenclature \cite{Vafa:2005ui}, one may refer to the RG trajectories within $\mathcal{S}_{\rm UV}$ as the ``landscape'' of theories consistent with quantum gravity while the ones being driven away from the fixed point as $k \rightarrow \infty$ lie in the ``swampland'' of effective field theories lacking a quantum gravity completion.

The UV-critical hypersurface in the vicinity of the fixed point is conveniently characterized by linearizing the flow. Expanding the $\beta$-functions around the fixed point, we obtain up to first order
\begin{equation}
\beta_{u_i}( \{u_j\})
\simeq
	\sum_j \mathbf{M}_{ij}	(u_j - u_j^*)
\text{.}
\end{equation}
The matrix $\mathbf{M}_{ij} = \partial \beta_{u_i} / \partial_{u_j}$ denotes the stability matrix.  Diagonalizing the stability matrix then allows us to write down the solution to the linearized flow in terms of the right-eigenvectors $V_I$ and eigenvalues of $\mathbf{M}$, satisfying $\mathbf{M} \, V_I = - \theta_I V_I$:
\begin{equation}
u_i(t)
=
u_i^*
+	C_I	\, V_{Ii} \,	\exp(-\theta_I t)
\text{.}
\end{equation}
The $\theta_I$ are referred to as the critical exponents. The numbers $C_I$ determine the initial conditions of the flow, and are a priori free parameters of the theory.
However, if $\theta_I <0$, the only way to end at the fixed point as $t\to \infty$ is if the corresponding $C_I$ vanishes. In this case, the eigendirection $V_I$ is called UV-irrelevant. Conversely, a positive critical exponent automatically runs into the fixed point. Thus the parameter $C_I$ is undetermined by asymptotic safety and has to be fixed by experiment. The corresponding eigendirection is  UV-relevant. The dimension of the UV-critical hypersurface is therefore given by the number of relevant directions of the fixed point.

The Wilsonian viewpoint on renormalization is perfectly suited to explain the problems encountered in the perturbative quantization of the Einstein-Hilbert action. Studying the flow of the Einstein-Hilbert action (see also Section \ref{sec:EinsteinHilbert}), we can calculate the RG flow of Newton's coupling $G$ using perturbation theory. We find that the point $G=0$ is indeed a fixed point. The critical exponent associated with $G$ is given by the mass-dimension of the coupling, $\theta = -2$, implying that the RG flow is repelled from the GFP in the UV. The observation that $G$ is actually non-zero then entails that the Einstein-Hilbert action is not part of the UV-critical hypersurface \emph{of the Gaussian fixed point} and thus not asymptotically free. 
In perturbation theory the resulting divergences may be cured by adding higher-order counterterms \cite{tHooft:1974toh,Goroff:1985th}; however for gravity it turns out that the number of required counterterms is infinite. This spoils the predictivity requirement, stating that the Einstein-Hilbert action results in a perturbatively non-renormalizable quantum field theory.

Weinberg \cite{Weinberg:1980gg} conjectured that renormalizability of gravity may be restored by the existence of a non-perturbative fixed point. The study of non-perturbative techniques gained momentum with the development of Functional Renormalization Group Equations (FRGEs) applicable to gravity \cite{Reuter:1996cp}
\begin{equation}\label{FRGE}
\partial_t	\Gamma_k[\phi]
=
\frac{1}{2}	\mathrm{STr} \left[	\left(	\Gamma_k^{(2)}[\phi] + R_k \right)^{-1}	\partial_t	R_k	\right]
\text{.}
\end{equation}
In this equation, $\Gamma^{(2)}[\phi]$ denotes the Hessian of $\Gamma_k$, and $\mathrm{STr}$ the trace over all fluctuations, including a minus sign for ghost and fermionic modes. The RG is implemented by a $k$-dependent regulator $R_k$, which suppresses modes of momentum $p \lesssim k$.

Using these non-perturbative techniques, substantial evidence has been found for the existence of a NGFP for gravity. Starting from the seminal work \cite{Reuter:1996cp} elaborate investigations have accumulated substantial evidence supporting this conjecture \cite{Niedermaier:2006wt,Codello:2008vh,Litim:2011cp,Reuter:2012id,Nagy:2012ef,PercacciBook,RSBook}.

\section{Renormalization group flow of $f(R)$-gravity}
\label{sec:EinsteinHilbert}

A well-studied approximation of the gravitational effective average action is the $f(R)$-truncation \cite{Codello:2007bd,Machado:2007ea,Codello:2008vh}
\begin{equation}\label{fRansatz}
\Gamma_k[g]
\simeq
\frac{1}{16\pi G_k}	\int	\dd^4 x	\sqrt{g}	\bar{f}_k(R)
+	\Gamma_{\text{gf}}	+ \Gamma_{\text{gh}}
\text{,}
\end{equation}
where $R$ denotes the Ricci scalar. The EAA also contains a suitable gauge fixing $\Gamma_{\text{gf}}$ and ghost terms $\Gamma_{\text{gh}}$. The running couplings in this truncation are Newton's coupling $G_k$ and those contained in the function $\bar{f}_k(R)$. The (Euclidean) Einstein-Hilbert action is obtained by setting
\begin{equation}\label{EHaction}
\bar{f}_k(R) = - R + 2 \Lambda_k \, . 
\end{equation}
In principle, the function $\bar{f}_k$ contains infinitely many couplings that need to be fixed, providing an excellent testing ground for the predictivity of Asymptotic Safety.

In order to define the functional variation of the EAA, we employ the background field method. The easiest way to implement this is by a linear split of the metric,
\begin{equation}
g_{\mu\nu}
=
\bar{g}_{\mu\nu}	+	h_{\mu\nu}
\text{,}
\end{equation}
where $\bar{g}_{\mu\nu}$ is a fixed but arbitrary background metric and $h_{\mu\nu}$ parameterizes the fluctuations with respect to this background. The scale-dependence of $G_k$ and $\bar{f}_k(R)$ is obtained by substituting the ansatz \eqref{fRansatz} into the FRGE \eqref{FRGE} and projecting the result on actions of the $f(R)$-type. This results in a partial differential equation governing the scale-dependence of $\bar f_k(R)$. Introducing the dimensionless quantities
\begin{equation}
r = k^{-2} R 
\text{,}\qquad
G_k
=
k^{-2}	g_k
\text{,}\qquad
\bar{f}_k(R)
=
k^2	f_k(r)
\text{,}
\end{equation}
the equation becomes autonomous and may serve as a generating equation for the 
 $\beta$-functions \eqref{beta1}. By now, several incarnations of such generating equations 
 have been constructed, differing in the choices for the gauge fixing and parameterization of the fluctuation field \cite{Codello:2007bd,Machado:2007ea,Benedetti:2012dx,Dietz:2012ic,Dietz:2013sba,Demmel:2015oqa,Ohta:2015efa,Ohta:2015fcu}. We illustrate some of the central properties arising from the projection of \eqref{fRansatz} to finite order polynomials in $\bar f_k(R)$ in the sequel. 

\subsection{Projecting onto the Einstein-Hilbert action}
As a first example, we discuss the simplest approximation to the full function $f(R)$, namely the Einstein-Hilbert truncation \eqref{EHaction}. 
Following the steps and choices made in \cite{Reuter:1996cp}, we arrive at the $\beta$-functions for the dimensionless Newton's coupling $g_k$ and the dimensionless cosmological constant $\lambda_k = k^{-2} \Lambda_k$
\begin{subequations} \label{eq:EHbetag}
	\begin{align}
	\partial_t g_k
	&=
	(2	+	\eta_N)	g_k \, , 
	\\
	\partial_t	\lambda_k
	&=
	L_1	g_k	+	g_k	\eta_N	L_2	-	(2-\eta_N)	\lambda_k \,
	\text{,}
	\end{align}
\end{subequations}
where $\eta_N = -	\partial_t	G_k / G_k$ is the anomalous dimension of Newton's coupling. The anomalous dimension has the form
\begin{equation}
\eta_N
=
\frac{B_1}{1	-	g_k	B_2}	g_k
\text{.}
\end{equation}
 The coefficients $L_i$ and $B_j$ are given by
\begin{subequations}
	\begin{align}
	B_1	
	=&	
	\frac{1}{12\pi} \bigg(
	-	16 \Phi^1_{1} \left(0\right)
	-	24 \Phi^2_{2} \left(0\right)
	% 	\\&
	+	20 \Phi^1_{1} \left(-2 \lambda_k	\right)
	-	72 \Phi^2_{2} \left(-2 \lambda_k	\right)
	\bigg)
	\\
	B_2
	=&
	-	\frac{1}{24\pi} \bigg(
	20 \tilde{\Phi}^1_{1}\left(-2 \lambda_k	\right)
	-	72 \tilde{\Phi}^2_{2}\left(-2 \lambda_k	\right)
	\bigg)
	\\
	L_1
	=&
	\frac{1}{4 \pi} \left(
	20 \Phi^1_{2} \left(-2 \lambda_k	\right)
	-	16 \Phi^1_{2} \left(0\right)
	\right)
	\\
	L_2
	=&
	-	\frac{1}{8\pi} 20 \tilde{\Phi}^1_{2}\left(-2 \lambda_k	\right)
	\text{.}
	\end{align}
\end{subequations}
When evaluated with a Litim-type regulator \cite{Litim:2001up}, the threshold functions $\Phi^p_n(w)$ and $\tilde{\Phi}^p_n(w)$ are particularly simple and read
\begin{equation}
\Phi^p_n(w)
=
\frac{1}{\Gamma(n+1)}	\frac{1}{(1+w)^p}
\qquad
\tilde{\Phi}^p_n(w)
=
\frac{1}{\Gamma(n+2)}	\frac{1}{(1+w)^p}
\text{.}
\end{equation}

\begin{figure}
	\centering
	\includegraphics[width=.75\textwidth]{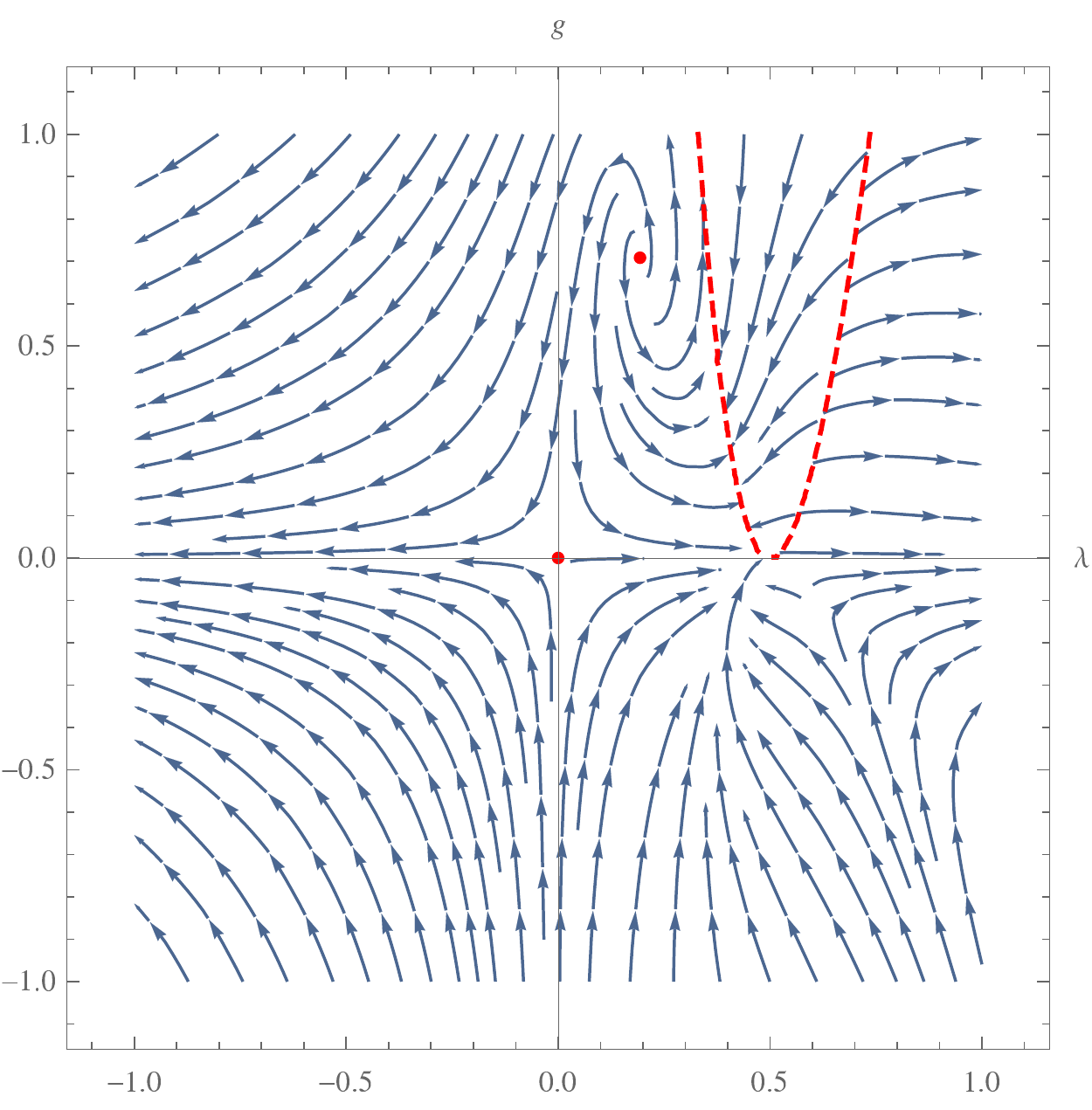}
	\caption{Overview of the Einstein-Hilbert phase diagram. The fixed points are indicated by red dots; flow lines by blue arrows. The red dashed line denotes the singularity in $\eta_N$. First obtained in \cite{Reuter:2001ag}.}
	\label{fig:EHphasediagram}
\end{figure}

The $\beta$-functions \eqref{eq:EHbetag} define a flow through the parameter space spanned by $g$ and $\lambda$. Fig.\ \ref{fig:EHphasediagram} gives an overview of the phase diagram.
For a large part, the flow is controlled by the interplay of two RG fixed points. The Gaussian fixed point (GFP) is located at
\begin{equation}
\lambda_*	=	g_*	=	0 \, . 
\end{equation}
We also find a non-Gaussian fixed point (NGFP) situated at
\begin{equation}
\lambda_* = 0.193
\text{,}
\qquad
g_*	=	0.707
\text{.}\label{eq:ehfp}
\end{equation}
Linearizing the flow around the fixed points, we find the canonical critical exponents for the GFP:
\begin{equation}
\text{GFP:}	\qquad	\theta_1	=	+2	\qquad \theta_2	=	-2
\text{.}
\end{equation}
The critical exponents of the NGFP are complex
\begin{equation}
\text{NGFP:}	\qquad	\theta_{1,2}	=	1.48	\pm	3.04\imath
\text{,}\label{eq:ehcritexp}
\end{equation}
The positive real part of $\theta_{1,2}$ indicates that the NGFP is UV-attractive. The non-zero imaginary part signals a spiraling behavior of the flow around the NGFP. The stability properties make the NGFP suitable for the Asymptotic Safety scenario.

Furthermore, from eq.\ \eqref{eq:EHbetag} it is clear that the locus $g=0$ is a zero for $\beta_g$. Therefore, the RG flow is unable to cross this line, as is also visible in Figure \ref{fig:EHphasediagram}. Since the measured value for Newton's constant is positive, this discards the lower half of the phase diagram as unphysical. Moreover, 
the anomalous dimension $\eta_N$ has a singularity at $g B_2 = 1$. This is visible in the phase diagram as the dashed red line where the flow changes direction. 

We are interested in studying  whether one of the RG trajectories generated by the $\beta$-functions are compatible with observations. Here we use that $G_k$ is measured to be of the order of $10^{-57}$ eV$^2$ at energy scales of $10^{-5}$ eV.
This indicates that $g_k = k^2 G_k$ is tiny on the measured energy scales. Thus, an RG trajectory reaches a classical regime close to $g=0$.

This motivates an expansion of the $\beta$-functions around $g=0$. From a practical viewpoint, it turns out to be convenient to rewrite the $\beta$-functions in terms of the new couplings
\begin{equation}
g_k
\text{,}
\qquad
\alpha_k	=	\lambda_k	g_k
\text{,}
\end{equation}
as this allows us to solve the expanded flow equations analytically.
We start with the $\beta$-function for $g$. Up to second order, this gives
\begin{equation}
\beta_g(g,\alpha)
\simeq
2g	-	\frac{7}{3\pi}	g^2	
\text{.}
\end{equation}
The analytic solution of $k \partial_k g_k = \beta_g(g)$ then reads
\begin{equation}\label{sol1}
g_k
=
\frac{6 \pi  g_{k_0} k^2}{7 g_{k_0} \left(k^2-k_0^2\right)+6 \pi  k_0^2}
\text{,}
\end{equation}
where $g_{k_0}$ is a integration constant specifying a particular RG trajectory.
For the dimensionful Newton's coupling eq.\ \eqref{sol1} entails
\begin{equation}
G_k
=
\frac{1}{1	+	\frac{7}{6\pi} G_{k_0} \left(k^2-k_0^2\right)}	G_{k_0}
\text{,}
\label{eq:1loopNewton}
\end{equation}
in accordance with Ref. \cite{Bonanno:2001xi}. We see that for $k^2 - k_0^2 \ll G_{k_0}$, Newton's coupling is approximately constant. Quantum corrections only occur as
\begin{equation}
k^2
\sim 
k_0^2 + 	\frac{6\pi}{7} G_{k_0}^{-1} \equiv k_G^2
\text{,}
\end{equation}
which is of the order of the Planck scale \eqref{MPlanck}. At energies above the Planck scale, Newton's coupling is driven to zero quadratically.

In order to determine the behavior of the cosmological constant, we expand the $\beta$-function for $\alpha_k$ up to second order in $g$
\begin{equation}
\beta_\alpha(g,\alpha)
\simeq
-	\frac{14}{3\pi}	\alpha	g	-	\frac{11}{3\pi}	g^2
\text{.}
\end{equation}
Plugging in the approximate solution for $g$ allows us to obtain an analytic expression for the $k$-dependence of $\alpha_k$ as well
\begin{equation}
\alpha_k
=
\frac{
	\alpha_{k_0}+\frac{11}{12\pi}	g_{k_0}^2 \left(1-\frac{k^4}{k_0^4}\right)
}{
	\left(1 +\frac{7}{6\pi}	g_{k_0} \left(k^2/k_0^2-1\right)\right)^2
}
\text{.}
\end{equation}
This gives the following IR behavior for $\Lambda_k$ \cite{Bonanno:2001xi}
\begin{equation}
\Lambda_k
=
\alpha_k	/	G_k
=
\frac{
	\Lambda_{k_0}	-	\frac{11}{12\pi}	G_{k_0}	\left(	k^4-k_0^4	\right)
}{
	1+\frac{7}{6\pi} G_{k_0} \left(k^2-k_0^2\right)
}
\text{.}
\label{eq:1loopCC}
\end{equation}
For sufficiently small values of $\Lambda_{k_0}$ this equation entails three scaling regimes for the cosmological constant. We see that the denominator changes the running of $\Lambda_k$ in the same way as $G_k$, starting at $k \sim G_{k_0}^{-1/2}$. Beyond this scale, the scale-dependence of $\Lambda_k$ is governed by the NGFP and the cosmological constant grows quadratically.
 However, the numerator also introduces a new scale where $\Lambda_k$ changes behavior, namely at
\begin{equation}
k^4	\sim	k_0^4	+ 	\frac{12\pi}{11}	\frac{\Lambda_{k_0}}{G_{k_0}} \equiv k_\Lambda^4
\text{.}
\end{equation}
Below this scale, the cosmological constant freezes out.  In the regime between $k_\Lambda$ and $k_G$, we see that the running of $\Lambda$ is proportional to $k^4$. In Section \ref{sec:cosmology} we discuss the possibility to fix the trajectory by asking that the infrared value of $\Lambda$ matches the observed value $\Lambda_{\text{obs}}$.

\subsection{Flows including an $R^{2}$-term}
The Einstein-Hilbert truncation can be extended to include an $R^2$-term in the $f(R)$-action.\footnote{For pioneering work in this direction see \cite{Lauscher:2002sq}.} 
From a theoretical perspective this extension is important since it gives rise to an additional free parameter. At the same time, it is of great phenomenological interest since 
it allows to generate an initial phase of accelerated expansion of the universe as in the Starobinsky's inflationary model \cite{Starobinsky:1980te,Starobinsky:1983zz}.

Explicitly, we parameterize the action by
\begin{equation}\label{r2trunc}
f_k(r)
=
2\lambda_k
-	r
+	b_k	r^2
\text{,}
\end{equation}
which supplements the Eintein-Hilbert action with the higher-derivative action \eqref{HDaction}.
We base our RG analysis on the generating equation derived in \cite{Machado:2007ea}. The resulting RG equations for $\lambda_k, g_k$ and $b_k$ possess a GFP situated at $\lambda=g=b=0$. In addition there is  a NGFP located at
\begin{equation}
\lambda_*
=
0.133
\text{,}\qquad
g_*
=
1.59
\text{,}\qquad
b_*
=
0.119
\text{.}
\end{equation}
Its critical exponents read
\begin{equation}
\theta_{1,2}
=
1.26 \pm 2.45\imath
\text{,}
\qquad
\theta_3
=
27.0
\text{.}
\end{equation}
The similarity of these values with the Einstein-Hilbert truncation, eqs. \eqref{eq:ehfp} and \eqref{eq:ehcritexp} suggest that this is actually the same NGFP seen in different projections of the RG flow. The three positive eigenvalues indicate that this fixed point has (at least) 
 three relevant directions.

\subsection{Higher-order truncations}
In order to further explore the predictivity of the NGFP higher order terms in the scalar curvature have to be included. At the level of polynomial $\bar{f}(R)$-approximations this has been done systematically up to order $R^6$ \cite{Codello:2007bd,Machado:2007ea}, $R^8$ \cite{Codello:2008vh}, $R^{35}$ \cite{Falls:2013bv,Falls:2014tra}, and recently $R^{70}$ \cite{Falls:2018ylp}. As key results, the corresponding analysis established that adding higher-derivative terms beyond $R^2$ does not give rise to additional relevant directions. Moreover, power-counting has been identified as a good ordering principle for judging the relevance of the higher-derivative term.\footnote{The system constructed by Ohta et.\ al.\ \cite{Ohta:2015efa,Ohta:2015fcu,Alkofer:2018fxj,deBrito:2018jxt,Alkofer:2018baq} uses a manifestly different gauge fixing and an exponential split of the metric fluctuations. The resulting NGFP comes with 2 relevant directions. A detailed analysis of the flow shows that this NGFP does not support a crossover to a semi-classical regime as displayed in Fig.\ \ref{fig:EHphasediagram} but sits on the other side of a singular locus. Thus it does not lend itself to the type of analysis discussed in the present work.}

As an illustration, we start from the generating equation derived in 
\cite{Machado:2007ea} and expand $f(r)$ up to third order:
\begin{equation}
f_k(r)
=
2\lambda_k
-	r
+	b_k r^2
+	c_k r^3
\text{.}
\end{equation}
Again one finds the projection of the NGFP on this 4-parameter space which is located at
\begin{equation}
\lambda_*=	0.132
\quad
g_*=	1.02
\quad
b_*=	0.0356
\quad
c_*=	-0.534
\text{,}
\end{equation}
and possesses critical exponents
\begin{equation}
\theta_{1,2}=	2.67	\pm	2.26	\imath
\qquad
\theta_3=	2.06
\qquad
\theta_4=	-4.42
\text{.}
\end{equation}
In addition to the three relevant directions encountered in the previous section, there is also one irrelevant direction. Asymptotic safety then implies one relation between the four couplings contained in the ansatz. By linearizing the flow around the fixed point, we can find an equation for the coupling $c$  in terms of the couplings $\lambda$, $g$ and $b$:
\begin{align}
c
&=
-0.575	+0.434 \lambda	-	0.0583 g	+	1.21 b
\text{.}
\end{align}
This relation, holding at very high energies, ensures that the corresponding RG trajectories sit in the UV-critical hypersurface of the NGFP. At lower energies, the flow must be integrated down to the desired energy scale $k$. This amounts to integrating a highly nonlinear flow, which must be executed numerically. Doing so gives a prediction for the operator $\frac{C_k}{16\pi G_k} \int R^3$, which can be tested against observational bounds. 

%-------------------------------------------------------------------------------
\subsection{The non-Gaussian fixed point beyond $f(R)$-truncations}
%-------------------------------------------------------------------------------
We close our discussion with the following remarks. Going beyond approximations built from functions of the scalar curvature $R$, the inclusion of a Weyl-squared term to the higher-derivative action \eqref{r2trunc} has been considered in \cite{Benedetti:2009rx,Benedetti:2009gn,Hamada:2017rvn}. In this case quantum corrections turn the associated dimensionless coupling in a irrelevant one, so that the enhancement of the approximation does not introduce additional free parameters. 
Moreover, supplementing the Einstein-Hilbert action by the perturbative two-loop counterterm found by Goroff and Sagnotti \cite{Goroff:1985th} showed that the NGFP also persists in this setting \cite{Gies:2016con}: in contrast to the perturbative quantization procedure, the new direction is irrelevant at the NGFP and does not introduce a free parameter.

Along a different path the function $\bar{f}_k(R)$ may be replaced by a function of the squared Ricci curvature $\bar{f}_k(R_{\mu\nu} R^{\mu\nu})$ or the Riemann tensor $\bar{f}_k(R_{\mu\nu\rho\sigma} R^{\mu\nu\rho\sigma})$. A first analysis \cite{Falls:2017lst} showed that polynomial expansions in these quantities also see three free parameters, in agreement with the $f(R)$-analysis.

%------------------------------------------------------------------------------
\section{Matching cosmological observations}\label{sec:cosmology}
%------------------------------------------------------------------------------
Following the analysis in \cite{Gubitosi:2018gsl}, we now fix the three free parameters appearing in the $f(R)$ approximation of Asymptotic Safety. The key idea is that, for a given $f(R)$ action, the specific RG trajectory realized in nature can be identified by providing   the measured values of the couplings at given energy scales as initial conditions. We first illustrate how this is done for the Einstein-Hilbert truncated action, where the observed values of the Newton's coupling and of the cosmological constant are sufficient to fully determine a viable RG trajectory (also see \cite{Reuter:2004nx} for earlier work). We then show that also the trajectory associated to the $R^{2}$-truncated action can be fixed if one assumes that the higher-curvature term is responsible for the accelerated expansion of the universe in the inflationary epoch. Interestingly, the predicted runnings of the Newton's coupling and of the cosmological constant in this last case do not differ much from the ones predicted by the Einstein-Hilbert truncation.

\subsection{Newton's coupling and cosmological constant}
Observational constraints on Newton's coupling and the cosmological constant refer to different scales as summarized in Table \ref{tab:cosmologicaldata}.
\begin{table}
	\renewcommand{\arraystretch}{1.5}
	\centering
	\begin{tabular}{ll}
		Energy scale (eV)	&	RG constraint
		\\\hline
		$k \simeq	k_{\text{lab}} = 10^{-5}$	&	$G_k \simeq G = 6.7 \times 10^{-57}$ eV$^{-2}$
		\\
		$k \simeq	k_{\text{Hub}} = 10^{-33}$	&	$\Lambda_k \simeq \Lambda = 4 \times 10^{-66}$ eV$^2$
		\\
		$k \simeq k_{\text{infl}} = 10^{22} $ & $B = - 6.7 \times 10^{-39}$ eV$^{-2}$
		\\\hline
	\end{tabular}
	\caption{Observational constraints on the parameters of the Einstein-Hilbert action parameters with corresponding energy scales (top two lines). Imposing that the effective average action in the $R^2$-approximation gives rise to Starobinsky-inflation yields the additional constraint shown in the bottom line. From \cite{Gubitosi:2018gsl}.}
	\label{tab:cosmologicaldata}
\end{table}
Since at the scales where the observations take place the dimensionless Newton's coupling is well within the perturbative regime, $g \ll 1$, we can use the approximated flow equations \eqref{eq:1loopNewton} and \eqref{eq:1loopCC} to extrapolate the RG flow to an initial point $(\alpha_{k_0},g_{k_0})$ that lies on a trajectory that satisfies the observed data. Starting from this point, the $\beta$-functions \eqref{eq:EHbetag} can be integrated numerically to obtain the full trajectory.

\begin{figure}
	\centering
	\includegraphics{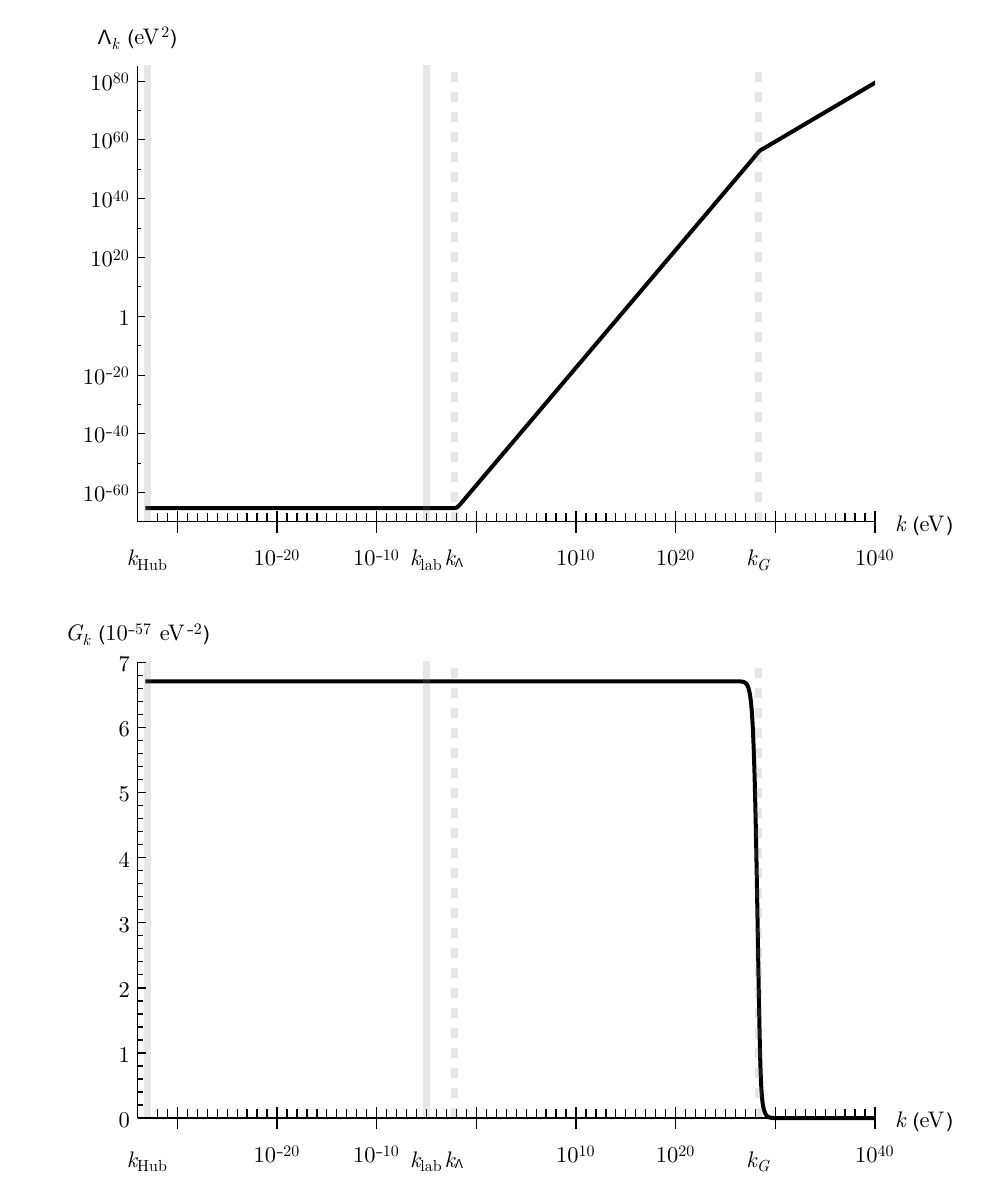}
	\caption{RG trajectory satisfying cosmological observations. Observational constraints are imposed at the laboratory scale $k_{\text{lab}}$ and Hubble scale $k_{\text{Hub}}$, denoted by the solid gray lines. The RG flow dynamically generates the scales $k_G$ and $k_\Lambda$, denoted by the dashed gray lines.}
	\label{fig:einsteinhilbert_phystraj}
\end{figure}

The  resulting integrated trajectory is shown in Fig.\ \ref{fig:einsteinhilbert_phystraj}. We observe that below the dynamical Planck scale $k_G$, Newton's coupling remains constant. Above this scale, the flow is controlled by the NGFP and $G_k = k^{-2} \, g_*$ decreases quadratically in $k$. The cosmological constant exhibits the three scaling regimes discussed in connection with eq.\ \eqref{eq:1loopCC}: for $k \lesssim k_\Lambda$ the cosmological constant is constant and agrees with the observed value. In the intermediate region $k_\Lambda \lesssim k \lesssim k_G$, $\Lambda_k \propto k^4$ while for $k \gtrsim k_G$ the flow is governed by the NGFP entailing that $\Lambda_k = k^2 \lambda_*$ increases quadratically with $k$. Notably, the increasing value of $\Lambda$ above $k_\Lambda$ is compatible with current  planetary and atomic observational constraints \cite{Martin:2012bt}. It could however affect primordial perturbations and be detectable in the cosmic microwave power spectrum.

%--------------------------------------------------------------------------------------
\subsection{Constraints on $R^{2}$-truncated action from  early universe observations}
\label{sect4.2}
%--------------------------------------------------------------------------------------

\begin{figure}
	\includegraphics{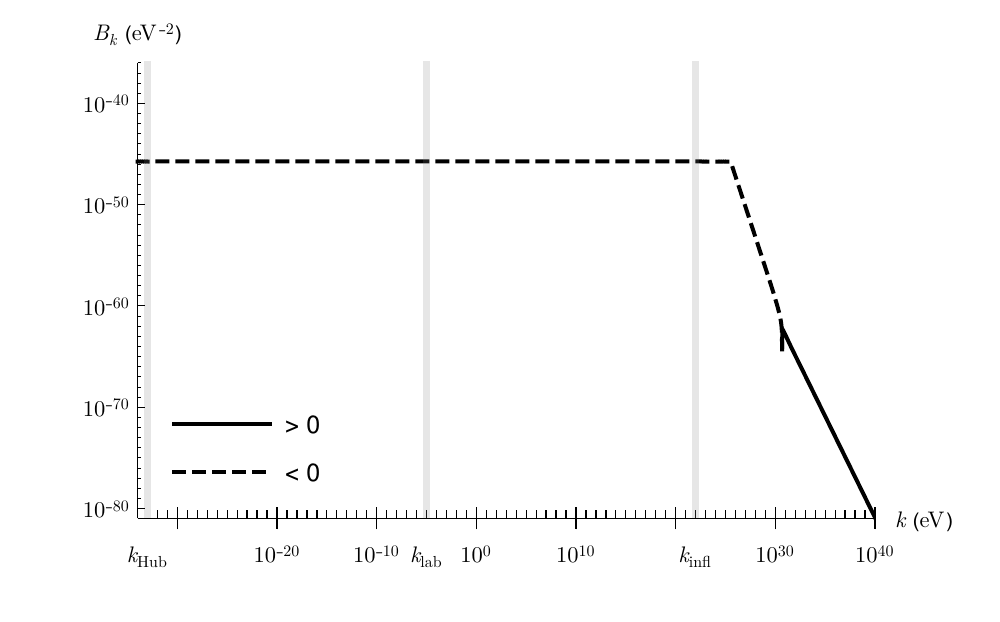}
	\caption{Running of the $R^2$-coupling $B_k$ satisfying observational constraint. The solid line denotes positive values of $B$, whereas the dashed line indicates negative $B$. The change of sign to a negative value of $B_k$ at the scale of inflation is essential for producing the correct inflationary dynamics \cite{Netto:2015cba}.}
	\label{fig:inflation_phystraj}
\end{figure}

When considering higher order curvature terms in the $f(R)$ truncated action, observational constraints on the Newton's coupling and the cosmological constant are insufficient to determine the physical trajectory completely. This is because there is one additional free parameter, given by the $R^{2}$ coupling. 
The additional initial condition can be fixed by assuming that this coupling is responsible for the initial era of inflationary expansion of the universe. 
Constraints on inflation are set by observations on the cosmic microwave background (CMB), as recently done by the PLANCK collaboration \cite{Ade:2015lrj}. The reported value for $B_k$ is \cite{Martin:2013tda,Copeland:2013vva}
\begin{equation}
M_P^2 \,	B_k
\simeq
-	1	\times 10^{9}
\text{,}
\end{equation}
where $B_k= k^{-2}b_k$ and $k$ should be taken to be at the scale of inflation, which is placed at $k_{\text{infl}} = 10^{22}$~eV.
Together with the observational constraints from Table \ref{tab:cosmologicaldata}, this provides three constraints for an RG trajectory associated to the $R^{2}$ truncated $f(R)$ action. 
In \cite{Gubitosi:2018gsl}, we have constructed the RG flow corresponding to these constraints. The flow of $B_k$ is shown in Fig.\ \ref{fig:inflation_phystraj}. Remarkably, he behavior of $\Lambda_k$ and $G_k$ is identical to the one found in the Einstein-Hilbert case, Fig.\ \ref{fig:einsteinhilbert_phystraj}, up to minor numerical corrections. 

\subsection{Constraints on higher-order truncations}
The cosmological analysis can be extended to include higher powers of $R$. Viable RG trajectories should now flow towards the NGFP in the UV, and yield an effective action compatible with observations at the scale of inflation. Since the Starobinsky model fits observational data with extremely high precision, there is an upper bound on the value of the dimensionful higher-order couplings.
Whether the RG flow of $f(R)$-gravity supports these conditions requires a numerical integration of the beta functions.

An interesting step in this direction has been taken in \cite{Liu:2018hno}. Motivated by the structure of the effective average action obtained from solving the generating equation at the NGFP \cite{Benedetti:2012dx,Dietz:2012ic,Dietz:2013sba,Demmel:2015oqa}, the higher order terms appearing in the polynomial $f(R)$ expansion have been resumed into a logarithm. The resulting refined Starobinsky model corresponds to\footnote{A cosmological analysis of similar models obtained from a renormalization group improvement of the Einstein-Hilbert action has been \cite{Bonanno:2012jy,Bonanno:2015fga}, also see \cite{Bonanno:2017pkg} for a review and further references.} 
\begin{equation}\label{refs}
\bar{f}_{k_{\text{infl}}}(R) = R + B \, R^2 \, (1 + A \ln(R/\mu^2)) \, ,
\end{equation}
where $B$ and $A$ are constants and $\mu$ an energy scale. For $A=0$ this model reduces to the Starobinsky-model discussed in Subsection \ref{sect4.2}. Increasing $A$ increases both the scalar spectral index $n_s$ and the tensor-to-scalar ratio $r$. Requiring consistency of $n_s$ with the Planck data permits tensor-to-scalar ratios up to $r \approx 0.01$ which allows to test this type of models with the next generation of CMB experiments.

An important prerequisite for a successful generalization of the $R^2$-scenario towards the inclusion of higher-order curvature terms is the continued existence of the GFP, responsible for a classical scaling regime. Depending on the precise implementation of the gauge-fixing and regularization procedure this feature is not guaranteed automatically (also see \cite{Falls:2018ylp} for a related discussion). It is realized by the flow equation obtained in the framework of geometric flows \cite{Demmel:2014hla} which thus constitutes a feasible starting point for such a investigation in the future.
%------------------------------------------------------
\section{Summary and concluding remarks}
\label{sec:conclusion}
%------------------------------------------------------
The asymptotic safety scenario \cite{Niedermaier:2006wt,Codello:2008vh,Litim:2011cp,Reuter:2012id,Nagy:2012ef} gives a fascinating perspective for
obtaining a quantum theory of the gravitational force valid up to arbitrarily short distance scales. In this construction the gravitational dynamics at trans-Planckian scales is controlled by a fixed point of the gravitational renormalization group flow (NGFP) corresponding to an interacting theory. As a consequence, all dimensionless quantities stay finite and physical processes are free from unphysical divergences at high energies. In addition the NGFP equips the construction with significant predictive power: the condition that an asymptotically safe theory flows into the NGFP at increasing energy 
fixes the infinite number of gravitational couplings in terms of a few relevant parameters. The latter must be taken from experimental input. At the present stage, it is conceivable that the free parameters of the theory can be linked to the value of the cosmological constant, Newton's coupling, and one four-derivative coupling as, e.g., the coupling encoding the strength of the $R^2$-interaction. While the identification of the complete set of free parameters is still ongoing, the observation that classical power counting still provides a good guiding principle for the relevance of an interaction \cite{Falls:2013bv,Falls:2014tra,Falls:2017lst,Eichhorn:2018ydy,Falls:2018ylp} lends strong support to the general arguments \cite{Benedetti:2013jk} that this set will be finite. Loosely speaking the interactive nature of the renormalization group fixed point gives rise to quantum corrections to the relevance of a coupling (in particular switching the marginal ones to either being relevant or irrelevant). At the same time, they are not strong enough to topple the hierarchy inferred from the classical analysis. 

As illustrated by Fig.\ \ref{fig:einsteinhilbert_phystraj} certain theories emanating from the NGFP undergo a crossover to a ``classical regime'' characterized by the dimensionful couplings becoming constant. The crossover scale between the fixed point and classical regime is set by the Planck scale $M_{\rm Pl}$. Similarly to $\Lambda_{\rm QCD}$ this scale is created dynamically and must be fixed by experimental observations. Solutions of the flow equations indicate that the observed value of the cosmological constant can be accommodated in the asymptotic safety construction: $\Lambda_{\rm obs}$ may be taken as an experimental input fixing one of the free parameters in the construction. While this viewpoint does not ``explain'' the tiny value of $\Lambda_{\rm obs}$ in a ``natural'' way, it ensures that the classical part of the solution extends up to cosmic scales. 

An interesting facet of asymptotic safety is the observation that the gravitational dynamics in the classical regime may actually consist of the Einstein-Hilbert action supplemented by additional interactions either built from higher-order curvature terms or non-local contributions. Owed to the smallness of the spacetime curvature, e.g., at solar system scales, there are no stringent bounds on the corresponding couplings \cite{Will:2014kxa}. 
At the same time quantum gravity induced interactions may play an important role in the very early universe or in explaining the accelerated expansion at late times without resorting to a cosmological constant. Combining asymptotic safety with the hypothesis that the phase of inflation occurring in the very early universe is actually driven by a modified gravitational dynamics leads to  
interesting cosmological predictions which may be tested experimentally.
For instance the $R^2$ coupling may be fixed from cosmic parameters extracted from the properties of fluctuations in the cosmic microwave background. 

Remarkably, albeit somewhat speculative, the gravitational renormalization group flow may also exhibit a mechanism for generating small values of the cosmological constant in a dynamical way. The cosmological constant naturally enters into the graviton propagator. For positive values of $\Lambda$, this gives rise to an instability of the propagator at low energy. It is conceivable that this instability may be cured by driving the cosmological constant to zero dynamically \cite{Wetterich:2017ixo}. Alternatively, the long-range nature of the gravitational interactions may lead to non-local terms in the effective action which mimick the dynamics resulting from a cosmological constant \cite{Maggiore:2014sia,Belgacem:2017cqo}. First evidence that such a scenario may indeed be realized has recently been provided in \cite{Knorr:2018kog} based on data \cite{Ambjorn:2008wc,Ambjorn:2016fbd} from Monte Carlo simulations within the Causal Dynamical Triangulation program \cite{Ambjorn:2012jv}. This interplay between quantum gravity and cosmological observations is 
predestined for new, exciting developments in the near future.

\bigskip
\noindent
{\bf Acknowledgements} F.S.\ thanks the organizers of the Workshop ``Naturalness, Hierarchy, and Fine Tuning'' for hospitality and putting together a very interesting scientific program. C.R.\ and F.S.\ acknowledge financial support by NWO through the FOM Vrije Programma 13VP12.
    
\bibliographystyle{spphys}       % APS-like style for physics

\end{document}